\documentclass[reprint,showpacs,amsmath,amssymb,aps,pre]{revtex4-1}
\usepackage{dcolumn}
\usepackage{bm}
\usepackage[dvipdfmx]{graphicx}
\usepackage[dvipdfmx]{hyperref}
\usepackage[mathlines]{lineno}
\usepackage{physics}

\begin{document}

\title{Topological phase transition in nonchiral Rice-Mele model with bond disorder}
\author{Kiminori Hattori, Kenyu Chikamori, Hayato Iizuka, and Ata Yamaguchi}
\affiliation{Department of Systems Innovation, Graduate School of Engineering Science, The University of Osaka, Toyonaka, Osaka 560-8531, Japan}
\date{\today}

\begin{abstract}	
The Rice-Mele model consists of a one-dimensional lattice with two sublattice sites in each unit cell subjected to a staggered sublattice potential.
The onsite potential constitutes a mass term that breaks chiral symmetry.
In this paper, we show that a topological phase transition is induced in this model by disordering intracell and intercell hopping energies unequally, by means of a symmetry-independent global invariant.
For small enough mass, the phase transition is accompanied by anomalous localization, which is accounted for in terms of geometric means of random variables.
The specific disorder strength at which anomalous localization occurs is independent of mass.
In contrast, the critical disorder strength at which the phase transition takes place decreases as mass increases, and eventually becomes invariable for large enough mass.
For large enough mass, we show that the phase transition is characterized by arithmetic means instead of geometric means.
\end{abstract}

\maketitle

\section{Introduction}
\label{sec:1}

In recent years, topological materials have attracted immense attention in condensed matter physics \cite{ref:1,ref:2,ref:3,ref:4,ref:5,ref:6}.
An important issue in the study of topological matter is to explore the topological phase transition in the presence of strong disorder.
In general, the nontrivial bulk topology is immune to disorder up to some characteristic strength, and eventually disappears when the disorder becomes strong enough.
Recently, this feature has been observed for the nonchiral Rice-Mele (RM) model \cite{ref:7} as well as the chiral Su-Schrieffer-Heeger (SSH) model \cite{ref:8,ref:9,ref:10} with diagonal onsite disorder.
For the former, the nontrivial bulk topology is identified by a real-space global invariant, which is valid even in the absence of chiral symmetry \cite{ref:7}.
The nontrivial topology is also supported by the bulk-boundary correspondence \cite{ref:7,ref:11}.
Conversely, it is known that strong enough disorder can induce the trivial-to-nontrivial phase transition.
The disorder-driven topological phase is known as the topological Anderson insulator (TAI) phase \cite{ref:12,ref:13,ref:14}.
In chiral-symmetric AIII and BDI classes, the TAI transition is theoretically predicted \cite{ref:15,ref:16} and experimentally observed \cite{ref:17} in the presence of off-diagonal disorder.
In these systems, the phase transition is accompanied by anomalous localization \cite{ref:18,ref:19}, which causes the exponential localization length to diverge.
This unusual localization property is known to be specific to off-diagonal disorder.
Accordingly, it may be natural to ask whether analogous phenomena occur in the nonchiral RM model after off-diagonal bond disorder is turned on.
We address this problem in the following parts of this paper.
	
The paper is organized as follows.
To be self-contained, we briefly review the topological invariants formulated in momentum space and in real space for the nonchiral RM model in Sec. \ref{sec:2}.
We employ the real-space formula to numerically identify the topological phase in the presence of bond disorder.
In Sec. \ref{sec:3}, we show that the topological phase transition is induced by disordering intracell and intercell hopping energies unequally.
The phase transition depends on a mass term consisting of the sublattice potential that breaks chiral symmetry.
For small enough mass, the phase transition is accompanied by anomalous localization, which is accounted for in terms of geometric means of random variables.
The critical disorder strength at which the phase transition takes place decreases as mass increases, and eventually becomes invariable for large enough mass.
On the other hand, mass is irrelevant to the specific disorder strength at which anomalous localization occurs.
For large enough mass, we show that the phase transition is characterized by arithmetic means instead of geometric means.
Finally, Sec. \ref{sec:4} provides a summary.

\section{Theoretical formulation}
\label{sec:2}

Here, we summarize the topological invariants used for identifying the topological phase in the RM model without chiral symmetry.
We also show that the topological phase transition is reflected in some bulk observables. 

\subsection{Winding number}
\label{sec:2A}

The RM model consists of a one-dimensional (1D) lattice with two sublattice sites in each unit cell subjected to a staggered sublattice potential.
This model is described by the tight-binding Hamiltonian $H = {\sum _{jj'\alpha \beta }} \ket{j,\alpha } H_{jj'}^{\alpha \beta } \bra{j',\beta }$, where $j \in \{ 1,2, \cdots ,N\} $ denotes the lattice position of the two-site unit cell, $\alpha ,\beta \in \{ A,B\} $ represents the sublattice degree of freedom in each cell, and $\ket{j,\alpha } $ is the basis ket at each site.
The matrix element is explicitly written as $H_{jj'}^{AB} = H_{j'j}^{BA} = v{\delta _{jj'}} + w{\delta _{j,j'+1}}$ and $H_{jj'}^{AA} = - H_{jj'}^{BB} = m{\delta _{jj'}}$, where $v$ ($w$) denotes the intracell (intercell) hopping energy, and $m$ describes the staggered sublattice potential.
A nonzero value of $m$ breaks chiral symmetry, and hence the RM model falls in the AI symmetry class.
If $m = 0$, the RM model is reduced to the SSH model with chiral symmetry in the BDI class.

The Hamiltonian is formulated in momentum space as $H(k) = {\mathbf{h}}(k) \cdot \boldsymbol{\sigma} $, where $\boldsymbol{\sigma} = ({\sigma _x},{\sigma _y},{\sigma _z})$ is the Pauli vector, and the 3D vector ${\mathbf{h}} = ({h_x},{h_y},{h_z})$ is composed of ${h_x} = v + w \cos k$, ${h_y} = w \sin k$, and ${h_z} = m$.
Consider the normalized vector ${\mathbf{\hat h}} = {\mathbf{h}}/h$, where $h = \sqrt {h_x^2 + h_y^2 + h_z^2} $ is the norm of $\mathbf{h}$.
The unit vector ${\mathbf{\hat h}}$ forms a closed loop on the Bloch sphere as $k$ goes across the 1D Brillouin zone.
The number of times ${\mathbf{\hat h}}$ passes around the $z$ axis, expressed as
\begin{equation}
\label{eq:1}
\nu  = \frac{1}{{2\pi }}\int_0^{2\pi } {dk\frac{{\partial \phi }}{{\partial k}}} ,
\end{equation}
defines the winding number, where $\phi = {\tan ^{-1}}({h_y}/{h_x})$ is the azimuthal angle.
For the present model, one finds $\nu = 1$ for $\left| v \right| < \left| w \right|$ and 0 for $\left| v \right| > \left| w \right|$ \cite{ref:7,ref:11}.
Note that $\nu$ is distinct from the Zak phase of each band in units of $2\pi$, which is not quantized unless $m=0$.
In other words, there exists no single-band invariant in the RM model.
However, summing over two bands, one finds that the total Zak phase amounts to the azimuthal winding number $\nu$ \cite{ref:7,ref:11}.
In this sense, $\nu$ is a global invariant for all bands.
This implies that the topological phase identified by $\nu$ is not protected by the gap separating two bands, and the phase transition is irrelevant to gap closure.

\subsection{Bulk observables}
\label{sec:2B}

The eigenequation $H(k) \ket{{u_r}(k)} = {\varepsilon _r}(k) \ket{{u_r}(k)} $ is solved to be ${\varepsilon _r} = rh$ and
\begin{equation*}
\ket{u_r} = \frac{1}{\sqrt {2h(h-r{h_z})}}{\pmqty{r({h_x}-i{h_y}) \\ h-r{h_z}}} ,
\end{equation*}
where $r = \pm 1$ denotes the band index.
It is implied from the above results that the following bulk observables reflect the topological phase transition at $\left| v \right| = \left| w \right|$.
The bandgap is expressed as $\Delta = 2{h_\mathrm{min}}$, where ${h_\mathrm{min}} = \sqrt {{{(\left| v \right| - \left| w \right|)}^2} + {m^2}} $ is the minimum norm occurring at $k = {k_\mathrm{min}} \in \{ 0,\pi \} $.
This means that the gap is reduced to its minimum $\Delta = 2\left| m \right|$ for $\left| v \right| = \left| w \right|$.
The same result is derived from the asymptotic form $h(k) = \sqrt {{v^2}{{(k - {k_\mathrm{min}})}^2} + {m^2}} $ for $\left| v \right| = \left| w \right|$.
Hence, $\left| m \right|$ corresponds to the mass at the transition point.
Recall that the phase transition irrelevant to gap closure does not contradict topological band theory since $\nu$ is not a single-band invariant but the global one that characterizes all energy states.
The sublattice polarization is given by ${P_r} = \mel{u_r}{\sigma _z}{u_r} = r{h_z}/h$ for the eigenstate $\ket{{u_r}} $.
This means that ${P_r} = r \operatorname{sgn} m$ for $\left| v \right| = \left| w \right|$ at the lower-band maximum and the upper-band minimum.
In other words, the gap minimization and the fully polarized band-edge states signify the topological phase transition.
Note that in a finite RM chain with open boundaries, fully polarized edge modes are formed at $\varepsilon = \pm m$ in the nontrivial phase \cite{ref:7,ref:11}.
This implies that the band-edge states are smoothly transformed into the edge modes.

\subsection{Real-space topological invariant}
\label{sec:2C}

It is easy to incorporate off-diagonal bond disorder into the RM model by considering $H_{jj'}^{AB} = H_{j'j}^{BA} = {v_j}{\delta _{jj'}} + {w_{j'}}{\delta _{j,j' + 1}}$, ${v_j} = v + \delta {v_j}$ and ${w_j} = w + \delta {w_j}$, where $\delta {v_j}$ and $\delta {w_j}$ are random variables each with zero mean.
We can use an equivalent but simpler representation such that $H = {\sum _{\left| {l-l'} \right| < 2}} \ket{l} {H_{ll'}} \bra{l'}$, where $l \in \{ 1,2, \cdots ,2N\} $ denotes the site index.
In the presence of bond disorder, the matrix element is given by ${H_{l,l+1}} = {H_{l+1,l}} = {v_j}$ for $l=2j-1$ and ${w_j}$ for $l=2j$, and ${H_{ll}} = m$ for $l=2j-1$ and $-m$ for $l=2j$.
The eigenequation $H \ket{{u_n}} = {\varepsilon _n} \ket{{u_n}} $ is numerically solvable, where $n \in \{ 1,2, \ldots ,2N\} $.
Sorting the eigenenergies $\{ {\varepsilon _1},{\varepsilon _2}, \cdots ,{\varepsilon _{2N}}\} $ in ascending order, $\ket{{u_{n,-}}} \in \{ \ket{{u_1}} , \cdots ,\ket{{u_N}} \} $ and $\ket{{u_{n,+}}} \in \{ \ket{{u_{N + 1}}} , \cdots ,\ket{{u_{2N}}} \} $ constitute the lower and upper bands, respectively.
In terms of the rectangular matrix ${({U_r})_{ln}} = \braket{l}{{u_{nr}}}$ and the momentum translation matrix $T = {e^{i\delta kX}}$, the real-space topological invariant is defined by
\begin{equation}
\label{eq:2}
Q = \frac{1}{\pi } \arg\sum\limits_r {\det U_r^\dag T U_r} ,
\end{equation}
where $\delta k = 2\pi /N$ is the momentum interval, ${X_{ll'}} = {x_l}{\delta _{ll'}}$ is the traceless position matrix, and $\arg z \in ( - \pi ,\pi ]$ is the principal value of the argument.
The invariant $Q$ is explicitly gauge independent, and is always quantized to be 0 or 1 regardless of symmetries \cite{ref:7}.
It should also be noted that $Q$ is a global index that involves all energy states.
The identity $Q = \nu $ is analytically shown in the absence of disorder \cite{ref:7}.

\begin{figure}
\centering
\includegraphics{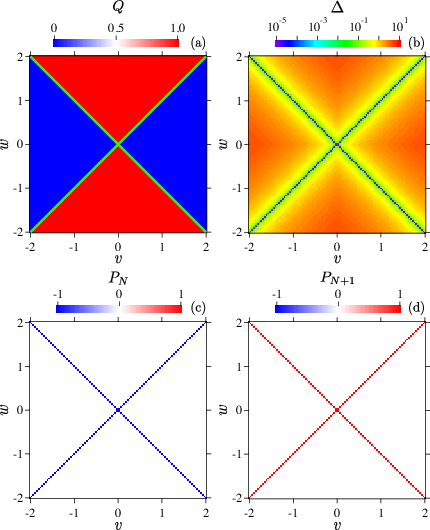}
\caption{(Color online) Maps of (a) $Q$, (b) $\Delta $, (c) ${P_N}$, and (d) ${P_{N+1}}$ in parameter space $(v,w)$ for an ordered periodic chain with $m = 5 \times {10^{-6}}$. In (a), the two crossed lines indicate the phase boundary at $\left| v \right| = \left| w \right|$.}
\label{fig:1}
\end{figure}

Figure \ref{fig:1} shows the real-space topological invariant $Q$ derived for a periodic ordered chain with $m = 5 \times {10^{-6}}$ and $N = 200$.
As seen in the figure, $Q$ is quantized to be 0 or 1, depending on relative magnitudes of $\left| v \right|$ and $\left| w \right|$.
At $\left| v \right| = \left| w \right|$, the topological phase transition takes place.
These features are retained for any values of $m$ in the absence of disorder, as expected from the identity $Q = \nu $.
In this figure, $Q$ is compared to $\Delta = {\varepsilon _{N+1}} - {\varepsilon _N}$, ${P_N} = \mel{u_N}{\sigma _z}{u_N} $ and ${P_{N+1}} = \mel{u_{N+1}}{\sigma _z}{u_{N+1}} $.
At the phase boundary, the gap is reduced to its minimum $\Delta = 2\left| m \right|$, and the two band-edge states $\ket{{u_N}} $ and $\ket{{u_{N+1}}} $ are fully polarized, i.e., ${P_N} = -1$ and ${P_{N+1}} = 1$.
These numerical results verify the above argument for bulk observables.

\section{Numerical calculation}
\label{sec:3}

Next, we proceed to numerical calculations for disordered systems.
In the calculations, we assume that the random variables $\delta {v_j} \in [-{W_v},{W_v}]$ and $\delta {w_j} \in [-{W_w},{W_w}]$ follow uniform distributions.
The numerical results shown below are derived for either ${W_v} = 0$ or ${W_w} = 0$.
Disorder averaging is performed over $M = {10^3} - {10^6}$  random configurations.
In what follows, the arithmetic mean of a quantity $O$ is denoted by $\ev{O} $.
The system size is typically $N = 200$.
The mass parameter is varied from $m = 0$ to 50.

\subsection{Phase transition and anomalous localization}
\label{sec:3A}

\begin{figure}
\centering
\includegraphics{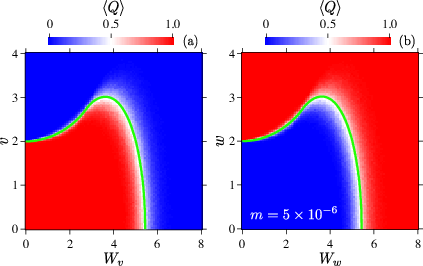}
\caption{(Color online) Maps of $\ev{Q} $ in parameter spaces (a) $({W_v},v)$ and (b) $({W_w},w)$ for $m = 5 \times {10^{-6}}$. The other parameters are fixed at $(w,{W_w}) = (2,0)$ in (a) and $(v,{W_v}) = (2,0)$ in (b). The solid lines inserted in these figures represent the solutions of Eq. (\ref{eq:3}).}
\label{fig:2}
\end{figure}

Figure \ref{fig:2} (a) shows $\ev{Q} $ as a function of ${W_v}$ and $v$ for $w = 2$ and $m = 5 \times {10^{-6}}$.
In the weak-disorder limit ${W_v} \to 0$, $\ev{Q} = 1$ for $\left| v \right| < 2$ and 0 for $\left| v \right| > 2$.
The phase boundary separating the nontrivial and trivial regimes varies with increasing the disorder strength ${W_v}$.
For strong enough disorder, the system is eventually in the trivial phase irrespective of $v$.
A contrasting behavior is seen in Fig. \ref{fig:2} (b), where $\ev{Q} $ is shown in the $({W_w},w)$  plane for $v = 2$.
In this case, strong enough disorder drives the system into the nontrivial phase regardless of $w$.
This is a TAI phase.
A local TAI behavior is also seen in Fig. \ref{fig:2} (a) for $v$ just exceeding 2, where weak disorder induces the trivial-to-nontrivial phase transition.
Similar phenomena are observed for chiral 1D systems in the AIII and BDI symmetry classes \cite{ref:12,ref:13,ref:14}.

\begin{figure}
\centering
\includegraphics{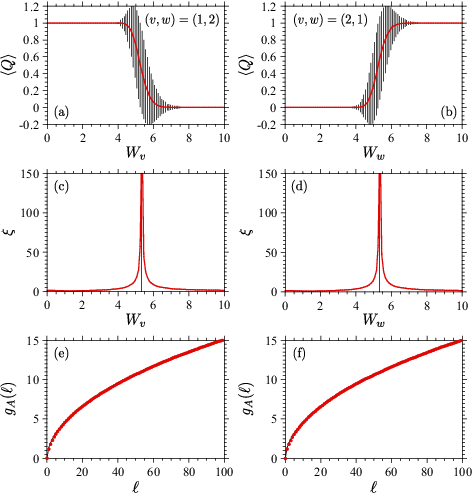}
\caption{(Color online) The upper two panels show $\ev{Q} $ as a function of (a) ${W_v}$ and (b) ${W_w}$. In these figures, error bars denote standard deviations. The middle two panels show $\xi $ derived at $\varepsilon = m$ as a function of (c) ${W_v}$ and (d) ${W_w}$. In these figures, vertical thin lines indicate the critical disorder strength ${W_v} = {W_w} = 5.34$ derived from Eq. (\ref{eq:3}). The lower two panels, (e) and (f), display ${g_A}(\ell )$ for $n = N+1$ computed at the critical disorder strength. In these figures, thin lines represent the power-law fit $ \propto {\ell ^{0.52}}$ to the data. In all of these calculations, we assumed $m = 5 \times {10^{-6}}$. The other parameters are fixed at $(v,w,{W_w}) = (1,2,0)$ in (a), (c) and (e), and $(v,w,{W_v}) = (2,1,0)$ in (b), (d) and (f).}
\label{fig:3}
\end{figure}

Figures \ref{fig:3} (a) and (b) explain disorder-inducing topological phase transitions more quantitatively for $(v,w) = (1,2)$ and $(2,1)$, respectively.
We see that $Q$ statistically fluctuates during the phase transition but becomes constant inside the trivial and nontrivial phases.
For chiral 1D systems, it is known that the topological phase transition of this type is accompanied by the divergence of the exponential localization length at $\varepsilon  = 0$ \cite{ref:12}.
In this study, the localization length $\xi $ at $\varepsilon = m$ has been evaluated by using the recursive Green's function formalism \cite{ref:7,ref:20,ref:21,ref:22}.
The derived results are shown in Figs. \ref{fig:3} (c) and (d).
As seen in the figures, $\xi $ tends to diverge at the critical point where the topological phase transition takes place.

The localization length is analytically derivable for the nonchiral RM model with bond disorder.
The relevant Schr{\"{o}}dinger equation is decomposed into
\begin{equation*}
(\varepsilon - m)\psi _j^A = {v_j}\psi _j^B + {w_{j-1}}\psi _{j-1}^B ,
\end{equation*}
\begin{equation*}
(\varepsilon + m)\psi _j^B = {v_j}\psi _j^A + {w_j}\psi _{j+1}^A ,
\end{equation*}
where $\psi _j^\alpha $ denotes the wavefunction at sublattice site $\alpha $ in cell $j$.
For $\varepsilon = m$, the set of equations is rearranged into the form $\ket{{\psi _{j+1}}} = {T_{j+1,j}}\ket{{\psi _j}} $, where $\ket{{\psi _j}} = {(\psi _j^A,\psi _j^B)^t}$ is the two-component column vector, and
\begin{equation*}
{T_{j+1,j}} = \pmqty{
  {-{v_j}/{w_j}} & {{2m}/{w_j}} \\ 
  0 & {-{w_j}/{v_{j+1}}} }
\end{equation*}
is the transfer matrix relating $\ket{{\psi _j}}$ and $\ket{{\psi _{j + 1}}}$.
Repeating this elemental relation leads to $\ket{{\psi _N}} = {T_{N1}}\ket{{\psi _1}} $ for the system of size $N$, where ${T_{N1}} = {T_{N,N-1}} \cdots {T_{32}}{T_{21}}$.
Since the matrix ${T_{j + 1,j}}$ is triangular, the diagonal elements of ${T_{N1}}$ are easily calculated to be $T_{N1}^{AA} = \prod _{j=1}^{N-1}(-{v_j}/{w_j})$ and $T_{N1}^{BB} = \prod _{j=1}^{N-1}(-{w_j}/{v_{j+1}})$.
They describe exponential localization since $\left| {\ev{\ln\left| {T_{N1}^{\alpha \alpha }} \right|} } \right| = (N-1)/\xi $ with ${\xi ^{-1}} = \left| {\ev{\ln\left| {{v_j}} \right|} - \ev{\ln\left| {{w_j}} \right|} } \right|$.
A similar result is obtained for $\varepsilon = - m$.
In this case, we have $\left| {\ev{\ln\left| {T_{1N}^{\alpha \alpha }} \right|} } \right| = (N-1)/\xi $.
Thus, the localization length $\xi $ at $\varepsilon = \pm m$ diverges under the condition
\begin{equation}
\label{eq:3}
\ev{\ln\left| {{v_j}} \right|} = \ev{\ln\left| {{w_j}} \right|} .
\end{equation}
Note that Eq. (\ref{eq:3}) is reduced to $\left| v \right| = \left| w \right|$ in the absence of disorder.
It is easy to find the analytic expressions for the two means appearing in Eq. (\ref{eq:3}).
For uniformly distributed random variables, we obtain $\ev{\ln\left| {{v_j}} \right|} = f(\left| v \right|,{W_v})$ and $\ev{\ln\left| {{w_j}} \right|} = f(\left| w \right|,{W_w})$, where
\begin{equation*}
f(x,y) = \frac{x}{{2y}}\ln\left| {\frac{{y+x}}{{y-x}}} \right| + \frac{1}{2}\ln\left| {{y^2}-{x^2}} \right| - 1 .
\end{equation*}
The present formula derived for the nonchiral model is independent of $m$, and reproduces the previous one for a chiral counterpart \cite{ref:15}.
The solution of Eq. (\ref{eq:3}) reasonably accounts for the phase transition at the critical point, as shown in Fig. \ref{fig:2}, and in Figs. \ref{fig:3} (c) and (d).
	
It may be instructive to reduce Eq. (\ref{eq:3}) to a simpler form.
Note that ${e^{\ev{\ln a} }}$ corresponds to the geometric mean ${\ev{a} _G} = {(\prod _{i = 1}^M{a_i})^{1/M}}$ for a set of positive numbers $\{ {a_1},{a_2}, \cdots ,{a_M}\} $.
Hence, Eq. (\ref{eq:3}) is reinterpreted as the equal geometric means ${\ev{\left| {{v_j}} \right|} _G} = {\ev{\left| {{w_j}} \right|} _G}$, or equivalently
\begin{equation}
\label{eq:4}
{\ev{v_j^2} _G} = {\ev{w_j^2} _G}
\end{equation}
for random disorder realizations.

Recall that $\ln\left| {T_{N1}^{AA}} \right| = \sum _{j=1}^{N-1}(\ln\left| {{v_j}} \right| - \ln\left| {{w_j}} \right|)$.
The term in parenthesis is a random variable with zero mean and finite variance at the critical point where $\xi \to \infty $.
Hence, $\ev{{{(\ln\left| {T_{N1}^{AA}} \right|)}^2}}  \propto N-1$.
This relation is reminiscent of an unbiased random walk in 1D, and implies that the relevant wavefunction decays as ${e^{- \lambda \sqrt \ell }}$ with distance $\ell $ from its maximum, where $\lambda $ is a positive scaling coefficient \cite{ref:18,ref:19}.
To see this, it is useful to calculate the correlation function ${g_\alpha }(\ell ) = \tfrac{1}{2}{\sum _ \pm }\ev{\ln\left| {\psi _{{j_\mathrm{max}}}^\alpha /\psi _{{j_\mathrm{max}} \pm \ell }^\alpha } \right|} $, where ${j_\mathrm{max}}$ denotes the cell at which $\left| {\psi _j^\alpha } \right|$ is maximal \cite{ref:19}.
It is expected that $g(\ell ) \propto \ell $ for ordinary exponential localization while $g(\ell ) \propto \sqrt \ell $ for anomalous localization.
The numerical results shown in Figs. \ref{fig:3} (e) and (f) demonstrate the latter at the critical point.

\subsection{Bulk observables and edge modes}
\label{sec:3B}

\begin{figure}
\centering
\includegraphics{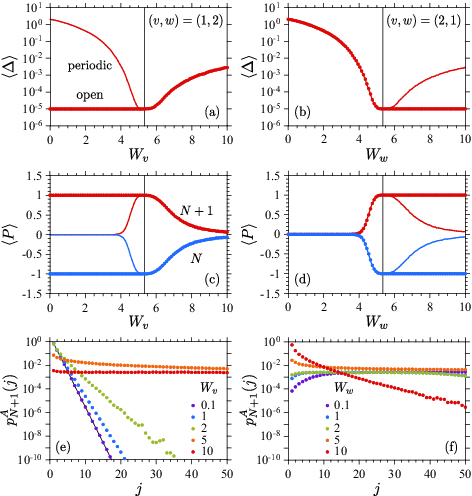}
\caption{(Color online) The upper two panels show $\ev{\Delta} $ as a function of (a) ${W_v}$ and (b) ${W_w}$. The middle two panels depict ${\ev{P_N}}$ and $\ev{P_{N+1}}$ as a function of (c) ${W_v}$ and (d) ${W_w}$. In these four figures, solid lines and dots represent the numerical results for periodic and open chains, respectively. Vertical thin lines indicate the critical disorder strength ${W_v} = {W_w} = 5.34$ derived from Eq. (\ref{eq:3}). The lower two panels, (e) and (f), display $p_{N+1}^A(j)$ for open chains with various disorder strengths indicated in the figures. In (e), a thin line shows the analytic probability density of the edge mode in the absence of disorder \cite{ref:7,ref:11}. In all of these calculations, we assumed $m = 5 \times {10^{-6}}$. The other parameters are fixed at $(v,w,{W_w}) = (1,2,0)$ in (a), (c) and (e), and $(v,w,{W_v}) = (2,1,0)$ in (b), (d) and (f). }
\label{fig:4}
\end{figure}

It is interesting to explore bulk observables during the phase transition.
Figures \ref{fig:4} (a) and (b) show the gap $\ev{\Delta} $ as a function of disorder strength.
It is seen in the figures that $\ev{\Delta} $ decreases down to $2\left| m \right|$ at the transition point for the disordered system with periodic boundaries.
This behavior is identical to that expected in the absence of disorder.
For open boundaries, $\ev{\Delta} $ is fixed to be $2\left| m \right|$ in the nontrivial phase.
This implies that two edge modes are formed at $\varepsilon = \pm m$.
Figures \ref{fig:4} (c) and (d) explain how the band-edge polarizations ${\ev{P_N}}$ and $\ev{P_{N+1}}$ vary with disorder strength.
As seen in the figures, $\ev{P} = \pm 1$ in the vicinity of the transition point for periodic boundaries, and in the nontrivial phase for open boundaries.
Again, the former coincides with the behavior exhibited by ordered systems, and the latter is ascribed to fully polarized edge modes.
These results exemplify that the topological phase transition is reflected in bulk observables even for disordered systems.
Figures \ref{fig:4} (e) and (f) display the disorder-averaged probability density $p_{N+1}^A(j) = \ev{{\left| {\braket{{j,A}}{{u_{N+1}}}} \right|}^2} $ in the region near the open boundary.
As expected, the edge mode emerges in the nontrivial phase and vanishes in the trivial phase.

\subsection{Dependence on sublattice potential}
\label{sec:3C}

\begin{figure}
\centering
\includegraphics{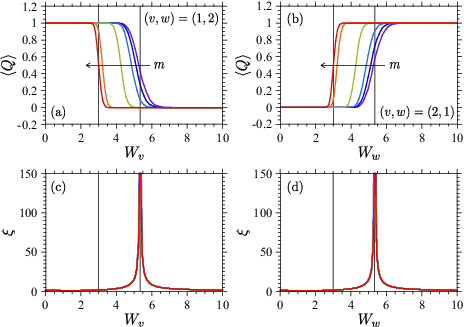}
\caption{(Color online) The upper two panels show $\ev{Q} $ as a function of (a) ${W_v}$ and (b) ${W_w}$ for $m$ varied as 0, 0.005, 0.05, 0.5, 5, and 50. The lower two panels summarize $\xi $ derived at $\varepsilon = m$ for various $m$'s as a function of (c) ${W_v}$ and (d) ${W_w}$. The other parameters are fixed at $(v,w,{W_w}) = (1,2,0)$ in (a) and (c), and $(v,w,{W_v}) = (2,1,0)$ in (b) and (d). Note that $\xi $ is identical for all values of $m$. In these figures, vertical thin lines indicate the critical disorder strengths ${W_v} = {W_w} = 5.34$ and 3 derived from Eqs. (\ref{eq:3}) and (\ref{eq:5}), respectively.}
\label{fig:5}
\end{figure}

\begin{figure}
\centering
\includegraphics{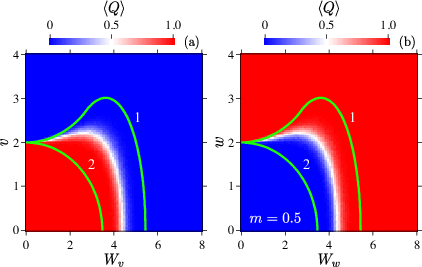}
\caption{(Color online) Maps of $\ev{Q} $ in parameter spaces (a) $({W_v},v)$ and (b) $({W_w},w)$ for $m = 0.5$. The other parameters are fixed at $(w,{W_w}) = (2,0)$ in (a) and $(v,{W_v}) = (2,0)$ in (b). The solid lines labeled with 1 and 2 represent the solutions of Eqs. (\ref{eq:3}) and (\ref{eq:5}), respectively.}
\label{fig:6}
\end{figure}

\begin{figure}
\centering
\includegraphics{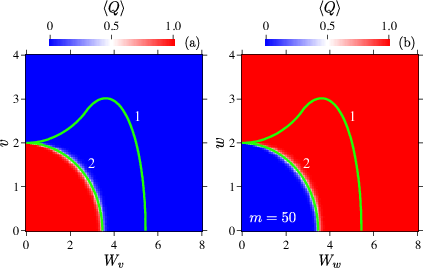}
\caption{(Color online) Maps of $\ev{Q} $ in parameter spaces (a) $({W_v},v)$ and (b) $({W_w},w)$ for $m = 50$. The other parameters are fixed at $(w,{W_w}) = (2,0)$ in (a) and $(v,{W_v}) = (2,0)$ in (b). The solid lines labeled with 1 and 2 represent the solutions of Eqs. (\ref{eq:3}) and (\ref{eq:5}), respectively.}
\label{fig:7}
\end{figure}

Thus far, we have assumed a relatively small mass.
A remaining question is how $m$ affects the topological properties of disordered systems.
Figures \ref{fig:5} (a) and (b) summarize the disorder-averaged topological invariant $\ev{Q} $ as a function of disorder strength for various values of $m$.
As seen in the figures, the critical disorder strength at which the phase transition takes place decreases as $m$ increases, and eventually becomes invariable for large enough mass.
For each $m$, the topological edge modes are observed in the nontrivial phase (not shown).
On the other hand, the localization length $\xi $ is fully independent of $m$, as shown in Figs. \ref{fig:5} (c) and (d).
The independence is consistent with the transfer-matrix formulation.
Thus, the phase transition is not always accompanied by anomalous localization, and is no longer described by Eq. (\ref{eq:3}) unless $m$ is small enough.
Figures \ref{fig:6} and \ref{fig:7} show the phase diagrams for $m = 0.5$ and 50, respectively.
As seen in the figures, the phase boundary sizably deviates from the solution of Eq. (\ref{eq:3}).

For large enough mass, we numerically find that the phase transition occurs when the condition
\begin{equation}
\label{eq:5}
\ev{v_j^2} = \ev{w_j^2}
\end{equation}
is met.
Equation (\ref{eq:5}) is a simple but reasonable relation since it is identical to $\left| v \right| = \left| w \right|$ in the absence of disorder.
Hence, it is natural to consider that the topological criticality in disordered systems is describable not only by Eq. (\ref{eq:3}) but also by Eq. (\ref{eq:5}).
The second-moments contained in Eq. (\ref{eq:5}) are explicitly expressed as $\ev{v_j^2} = {v^2} + W_v^2/3$ and $\ev{w_j^2} = {w^2} + W_w^2/3$ for uniformly distributed random variables.
As shown in Figs. \ref{fig:5} and \ref{fig:7}, the phase transition is precisely accounted for by Eq. (\ref{eq:5}) instead of Eq. (\ref{eq:3}) for large enough mass.
In addition, the phase transition fulfilling Eq. (\ref{eq:5}) does not contradict the bulk properties observed for $m = 50$ (not shown).
These observations support the implied criticality.
To summarize, the topological phase transition in the presence of bond disorder is characterized by the geometric means, Eq. (\ref{eq:3}) or (\ref{eq:4}), in the limit of $m \to 0$, whereas it is described by the arithmetic means, Eq. (\ref{eq:5}), in the limit of $\left| m \right| \to \infty $.

\section{Summary}
\label{sec:4}

We have investigated topological properties of the RM model with off-diagonal bond disorder.
The model includes a mass term that breaks chiral symmetry, and random hopping energies that violate translational invariance.
To numerically identify the topological phase in this model, we employ the real-space global invariant $Q$, which is quantized to be 0 or 1 regardless of symmetries.
Numerical results show that the topological phase transition is induced by disordering intracell and intercell hopping energies unequally.
The phase transition is also implied from bulk observables and edge modes.
For small enough mass, the phase transition is accompanied by anomalous localization, which is accounted for in terms of geometric means of random variables.
The critical disorder strength at which the phase transition takes place decreases as mass increases, and eventually becomes invariable for large enough mass.
On the other hand, mass is irrelevant to the specific disorder strength at which anomalous localization occurs.
For large enough mass, we show that the phase transition is characterized by arithmetic means instead of geometric means.
It is expected that the predicted phase transition could be experimentally tested on optical lattices.

\bibliography{ref}
 
\end{document}